\documentclass[aps,pre,twocolumn,showpacs]{revtex4}
\usepackage{graphicx}
\usepackage{epstopdf}
\usepackage{indentfirst}
\usepackage{braket}
\usepackage{float}
\usepackage{CJK}
\usepackage{esint}
\usepackage{color}
\usepackage{subfigure}
\usepackage{amsfonts}

\def\be{\begin{equation}}
\def\ee{\end{equation}}
\def\ba{\begin{eqnarray}}
\def\ea{\end{eqnarray}}

\begin{document}
\begin{CJK*}{UTF8}{gbsn}
\title{Equilibration of quantum chaotic systems}
\author{Quntao Zhuang(庄群韬)}
\affiliation{International Center for Quantum Materials, Peking University, Beijing 100871,China}
\author{Biao Wu(吴飙)}
\affiliation{International Center for Quantum Materials, Peking University,  Beijing 100871,China}
\affiliation{Collaborative Innovation Center of Quantum Matter, Beijing, China}
\date{\today}
\pacs{05.30.-d,05.45.Mt,03.65.-w}

\begin{abstract}
Quantum ergordic theorem for a large class of quantum systems was proved
by von Neumann [Z. Phys. {\bf 57}, 30 (1929)] and  again by
Reimann [Phys. Rev. Lett. {\bf 101}, 190403 (2008)] in a more practical and well-defined form.
However, it is not clear whether the theorem applies to quantum chaotic systems.
With the rigorous proof still elusive, we illustrate and verify this theorem  
for quantum chaotic systems with examples.
Our numerical results show that a quantum chaotic system with an initial low-entropy state
will dynamically relax to a high-entropy state and reach equilibrium. The quantum
equilibrium state reached after dynamical relaxation bears a remarkable resemblance
to the  classical micro-canonical ensemble.  However, the fluctuations around
equilibrium are distinct: the quantum fluctuations are exponential while the classical
fluctuations are Gaussian.
\end{abstract}
\maketitle
\end{CJK*}
\section{Introduction}
Boltzmann pondered on how to understand thermodynamics with the Newton's equations;
his answer to this question along with Gibbs' theory have become the foundation of classical
statistical mechanics ~\cite{huang_huang_1987,landau}. After quantum mechanics been
fully formulated, many giants in physics discussed a similar issue:  how to understand
thermodynamics with the  Schr\"odinger equation~\cite{eS,eV}.
In a 1929 paper, von Neumann provided an answer to this question by proving
two inequalities  ``in full rigor and without disorder assumptions"~\cite{eV}.
These two inequalities, which he called the quantum ergordic theorem and
the quantum H-theorem,  respectively,   laid down
a foundation for quantum statistical mechanics. However, this work has been largely
forgotten and apparently have never been mentioned in any modern textbook
on quantum statistical mechanics~\cite{huang_huang_1987,landau}.
There are discussions on why this work had almost been forgotten~\cite{goldstein}. In our opinion, one of the likely reasons is that  von Neumann introduced a 
rather unfamiliar concept, {\it macroscopic operators},
to prove his theorems.   It appears very hard to compute  these commuting
macroscopic operators, and related variables such as an entropy defined for a pure quantum
state~\cite{eV}.

Recently there have been renewed interests on the foundation of quantum statistical mechanics~\cite{zhuang_equilibration_2012,gemmer_quantum_2009,rigol_thermalization_2008,biroli_effect_2010,cassidy_generalized_2011,deutsch_quantum_1991,srednicki_chaos_1994,rigol_alternatives_2012,short_equilibration_2011,short_quantum_2012,popescu_entanglement_2006,popescu_foundations_2005,
linden_quantum_2009,cho_emergence_2010,ikeda_eigenstate_2011,rigol_relaxation_2007,reimann_foundation_2008,
reimann_equilibration_2012,xiong_universal_2011,goldstein_canonical_2006,suncp,wang,santos_weak_2012,reimann_typicality_2007,yukav_equilibration_2011, yukav_decoherence_2012, yukav_nonequilibrium_2011,riera_thermalization_2012,gogolin_absence_2011,christian_pure_2010,Emerson2013,Fine1,Fine2,Fine3,borgonovi_chaos_1998,flambaum_towards_1996} perhaps due to the remarkable progress in experimental realization of coherent quantum systems~\cite{Davis_BEC_1995,shin_distillation_2004,yao_prb_2006,zhao_anomalous_2011,huang_observation_2011,yao_nature}. An important result achieved is an inequality proved by
Reimann~\cite{reimann_foundation_2008,reimann_equilibration_2012} and later modified
by Short \it {et al.}\rm \cite{short_equilibration_2011,short_quantum_2012}. This
inequality can be regarded as a different version of von Neumann's quantum ergodic theorem.
The advantage of this new inequality is that every variable involved is well known and 
can be computed.  For this reason, when we discuss quantum ergodic theorem, we refer to
the inequality proved by Reimann unless stated otherwise.

According to the quantum ergodic theorem, an isolated quantum system starting
with a far-from-equilibrium state will relax dynamically to an equilibrium state and
stay there with very small fluctuations for almost all the time.  To be more specific,
for an isolated quantum system described by the wave function $\ket{\psi(t)}$,
it will relax to the following equilibrium state 
\be
\rho_\infty=\sum_k |c_k|^2\ket{E_k}\bra{E_k}\,,
\ee
where $\ket{E_k}$ is the energy eigenstate of the system and $c_k$'s are the 
expansion coefficients of $\ket{\psi(t)}$ in term of these eigenstates.  The density matrix 
$\rho_\infty$ is regarded as the micro-canonical ensemble by von Neumann~\cite{eV}. 
It is different from the usual micro-canonical ensemble found in textbooks
~\cite{huang_huang_1987,landau}, where the coefficients $c_k$'s take 
an identical value within a narrow energy shell.  As $|c_k|$'s
do not change with time, the micro-canonical density matrix $\rho_\infty$ is completely
determined by the initial condition. By utilizing this fact and the supposition principle,
we were able to predict a new quantum state which is at equilibrium with 
multiple temperatures, challenging the conventional
wisdom that an equilibrium state has only one temperature~\cite{zhuang_equilibration_2012}.

The quantum ergodic theorem holds only for quantum systems with no degenerate
energy gaps. Mathematically, this condition   is 
expressed as~\cite{eV,reimann_foundation_2008,reimann_equilibration_2012}
\begin{equation}
E_k-E_l=E_m-E_n\Rightarrow\left\{
\begin{array}{c}
E_k=E_l~{\rm and}~E_m=E_n\\
{\rm or}\\
E_k=E_m~{\rm and}~E_l=E_n
\end{array}\right.\,.
\label{condition}
\end{equation}
However, it is not clear at all how this condition of non-degenerate energy gap is related to
the familiar classification of quantum systems by their integrability. For a general integrable system, this condition is not satisfied as  quantum integrable systems  have the Poisson distribution of energy level spacing~\cite{introduction}, which imply 
the existence of many degenerate
eigen-energies.  However, there are plenty of examples of integrable systems, which have no
energy degeneracy at all. In fact, there are already reports of dynamical relaxation in integrable
systems~\cite{Santos,Santos2}.

The case for quantum chaotic systems is more complicated. As is well known,  quantum chaotic systems have the Wigner distribution of energy level spacing~\cite{introduction}. The Wigner distribution has two prominent features: zero probability at zero energy level spacing and a peak at a finite energy level spacing. The former feature means that there is little degeneracy,
which is favorite for the condition (\ref{condition}) being satisfied.
The latter implies that there are large number of energy level spacings 
around the peak value, which is clearly unfavorite for the condition (\ref{condition}) 
being satisfied. As a result,  it is not clear at all in the sense of mathematical 
rigor  whether the quantum ergodic theorem and H-theorem hold for 
quantum chaotic systems or not.  We led by intuition tend  to  believe 
that  the quantum ergodic theorem and  H-theorem should hold for quantum 
chaotic systems. Von Neumann believed
that his theorems should hold when the condition (\ref{condition}) is violated by
``infrequent exceptions"~\cite{eV}. This belief is now confirmed by 
Short and Farrelly~\cite{short_quantum_2012}.  Despite this progress, it is still not clear
how these mathematical conditions are related to the integrability of a system. 
The main purpose of this paper is to demonstrate
that the quantum ergordic theorem applies to quantum chaotic systems 
numerically with examples.

In this paper we study two quantum chaotic systems,
the ripple billiard~\cite{li_quantum_2002,xiong_universal_2011} and  the 
Henon-Heiles system~\cite{henon,zhuang_equilibration_2012}.
Our numerical simulation shows that both systems will indeed dynamically relax to
an equilibrium state where the overall features of the wave function no longer change.
For the ripple billiard system, where the successive energy-eigenstates can be computed,
the quantum ergodic inequality can be verified directly. In addition, we define entropies
for pure quantum states in the spirit of von Neumann as it is not clear  how to compute
the entropy defined for a pure quantum state by him. We find that these entropies will
approach maximized values,  offering another indication  that the system is indeed
equilibrating dynamically.

We have also analyzed  the properties of the equilibrium state reached after the dynamical relaxation in quantum chaotic systems. We  find an interesting correspondence between the quantum equilibrium state and the classical micro-canonical ensemble. We discuss 
the underlying mechanism with the correspondence between the quantum and the 
classical Liouville equations. At the end, we consider the statistical properties of 
fluctuations in quantum chaotic systems and find a distinction between the 
distributions of quantum fluctuations and the classical fluctuations.

This paper is organized as follows.   In Section \uppercase\expandafter{\romannumeral2} we introduce
the two  quantum chaotic systems, ripple billiard and Henon-Heiles system. In Section \uppercase\expandafter{\romannumeral3} we study the dynamical equilibration of quantum chaotic systems. In Section \uppercase\expandafter{\romannumeral4} we numerically verify the quantum ergodic theorem  in the ripple billiard system. In Section \uppercase\expandafter{\romannumeral5} we discuss the quantum-classical correspondence for the equilibrium states in detail. In Section \uppercase\expandafter{\romannumeral6} we discuss the fluctuation properties of the quantum systems and their corresponding classical systems. Finally in Section \uppercase\expandafter{\romannumeral7} we discuss the implications for many-body cases and summarize our results.

\begin{figure}
\centering
\subfigure[]{
\label{contourH} 
\includegraphics[width=0.21\textwidth ]{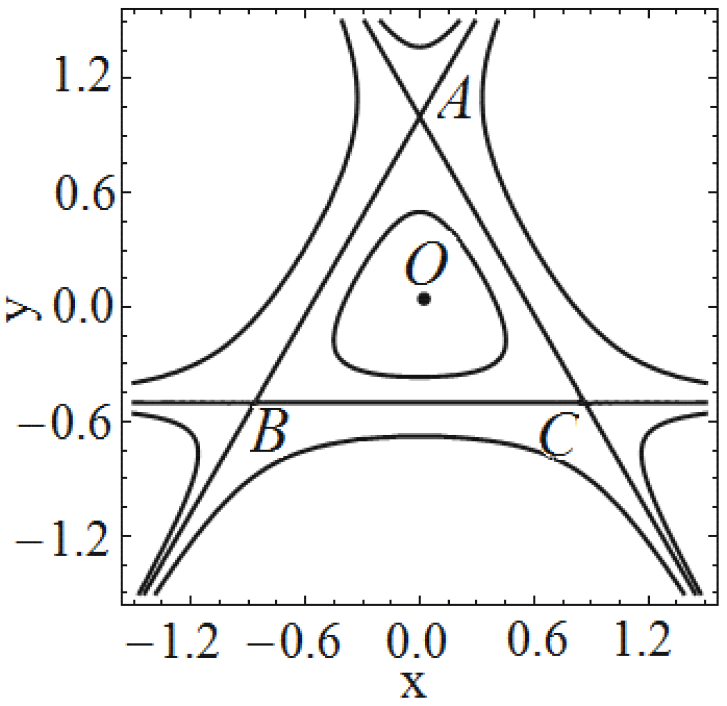}}
\centering
\subfigure[]{
\label{contourR} 
\includegraphics[width=0.23\textwidth]{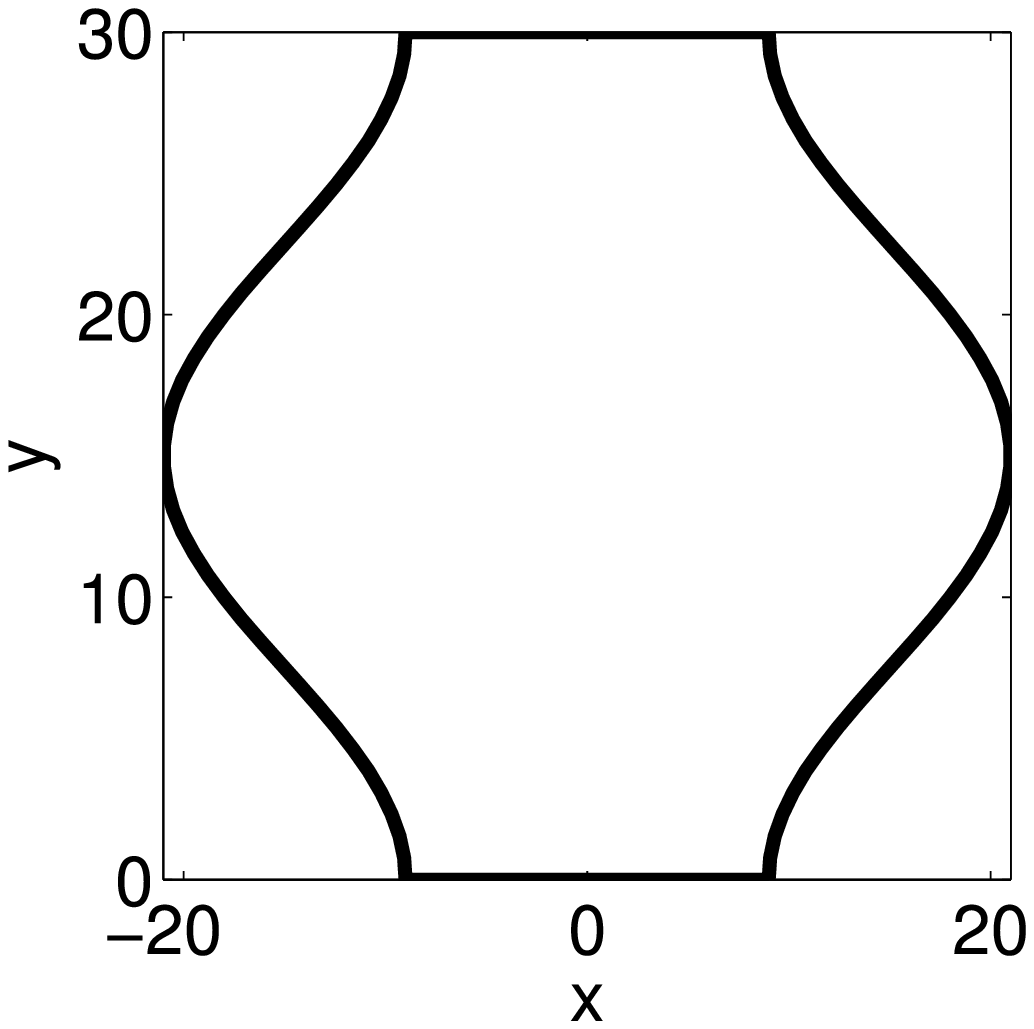}}
\caption{(a) Energy contours of the Henon-Heiles potential $V(x,y)$. The thick
solid lines are the contours  for $V/V_c=1/2,1,2$ from inside to outside.
The unit of axis is $r_c$. $A$, $B$ and $C$ are three saddle points and $O$ is the stable point.
(b) The shape of ripple billiard.
}
\end{figure}
\section{Models}

We focus on single particle quantum chaotic systems. There are two main reasons
for this choice.  First, single particle quantum systems are much less challenging numerically
and easier to analyze.  Second, according to random matrix theory,
the statistics of energy level spacings of quantum chaotic systems only depends on the type of
matrix of the system when its Hamiltonian is expressed in  an orthonormal basis~\cite{introduction}.
This property has nothing to do with whether the system is single particle or many body.
According to the condition (\ref{condition}),  the property of the eigen-energy spectrum
is the most important factor determining whether the quantum system will equilibrate or not.
There are properties that exist only in many-body systems, for example, the correlations~\cite{Flambaum}.
So far, no one has shown that the correlation plays any essential role in the equilibration process.
We also emphasize that we here consider isolated quantum systems and do not
consider the dynamics of quantum systems under external driving~\cite{Wilkinson}.
It is well known that a classical system will behave very differently under different drivings.
This feature seems to be shared by quantum systems~\cite{Wilkinson}.

A single particle in a two-dimensional chaotic potential is described by the Hamiltonian
\be
H={p^2}/{2m}+V(x,y)\,.
\ee
We choose the Henon-Heiles potential~\cite{henon} and the ripple billiard~\cite{li_quantum_2002}
as two examples for our study.
The Henon-Heiles potential is given by $V(x,y)= \frac{U}{2}(x^2+y^2)+\lambda (x^2y-\frac{y^3}{3})$. The energy contour of Henon-Heiles potential is shown in Fig. \ref{contourH}; it has four special points: one stable point O$(0,0)$, three saddle points A$(0,r_c)$, B$(-\frac{\sqrt{3}}{2}r_c,-\frac{1}{2}r_c)$, C$(\frac{\sqrt{3}}{2}r_c,-\frac{1}{2}r_c)$, where  $r_c\equiv\frac{U}{\lambda}$.
The classical orbits in the Henon-Heiles potential are chaotic when the energy is above $V_c/2$
and approach fully chaotic when the energy is close to $V_c$ with
$V_c\equiv\frac{U^3}{6\lambda^2}$. For later use, we set  $p_0=\sqrt{2mV_c}$.

Billiard systems are a two dimensional area surrounded by infinite potential walls at the edges.
For the ripple billiard~\cite{li_quantum_2002},  as shown in Fig. \ref{contourR}, the left and right edges are described by functions $x=\mp[b-a\cos(\pi y/b)]$ and the up and down edges are
two straight lines at $y=2b$ and $y=0$. The two geometrical parameters $a$, $b$ control the shape of 
the billiard .  When $a=0$ the ripple billiard is a square with width $2b$.  As $a$ increases from zero, it changes from an integrable system to a mixed, then to a fully chaotic system. It becomes mixed again when $a$ becomes very large. We have chosen for our computation $a=6,~b=15$, which corresponds to a fully chaotic case.

The two systems are chosen because each of them has its own advantages. For
the ripple billiard system, its successive eigenstates from the ground state up to the
3000th excited state can be computed numerically with great  precision~\cite{li_quantum_2002}.
To the best of our knowledge, for all other studied quantum chaotic systems, the high-energy eigenstates can
only be computed selectively~\cite{li_statistical_1994}.
The famous Henon-Heiles system is chosen to bring our study beyond billiards where $|p|$
is a constant of motion,  allowing us to gain more insights into general systems. 
In addition, we note that the Henon-Heiles system is beyond what is considered by 
von Neumann and others as it  has no bound eigenstates mathematically. However, 
its resonant states can be regarded as bound states in our numerically studies, where
the dynamical evolution lasts for a finite time and 
hard-wall boundaries are imposed at distance~\cite{hh_bound}. 

\section{Dynamical equilibration}
We numerically study the wave packet dynamics with the Schr\"odinger equation for
these two systems. In our numerical simulation we set $m=\frac{1}{2}$, $\hbar=1$.
The initial states are highly localized moving Gaussian wave packet for both systems,
\be
\psi(\vec{r},t=0)=\frac{\alpha}{\sqrt{\pi}}\exp(-\frac{1}{2}\alpha^2(\vec{r}-\vec{r_{i}})^2)\exp(i\vec{p_{i}}\cdot \vec{r}/\hbar)
\label{initial_condition}
\ee
with $1/\alpha=3r_c/40 $, $\vec{r}_{i}=(0.3,0)r_c$, and $\vec{p}_{i}=
\sqrt{7/10}(\cos10^\circ, \sin10^\circ)\,p_0$
for the Henon-Heiles system; $1/\alpha=a/6$, $\vec{r}_{i}=(0,0)$, and $\vec{p}_{i}=
(5,0)$ for the ripple billiard system. In both systems, a classical particle 
with the above initial position
$\vec{r}_{i}$ and momentum $\vec{p}_{i}$
has a fully chaotic orbit, which we confirmed  by computing  the Poincare section.
\begin{figure}
\includegraphics[width=0.48\textwidth ]{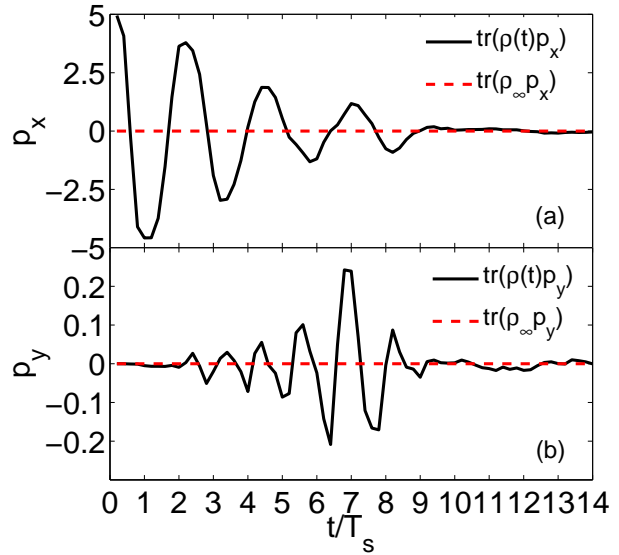}
\caption{Time evolution of (a) $\bra{\psi}P_x\ket{\psi}$ and (b) $\bra{\psi}P_y\ket{\psi}$
in the ripple billiard. The red lines are $\mbox{tr}(\rho_\infty p_x)$ and $\mbox{tr}(\rho_\infty p_x)$. The initial condition of $\vec{P}$ is (5,0). $T_s\equiv\frac{2(a+b)}{|p_i|/m}$.}
\label{momentum}
\end{figure}

The subsequent dynamical evolutions of these two Gaussian wave packets are computed.
For the ripple billiard, the dynamical evolution can be found in Ref.~\cite{xiong_universal_2011}; for
the Henon-Heiles system, the evolution is illustrated  in Ref.~\cite{zhuang_equilibration_2012}.
Both evolutions are very similar to each other. Here is a brief description: the smooth Gaussian
wave packet starts to spread out and gets diffracted by the potentials; the interference between
diffracted waves begins to make the wave packet appear more and more irregular; eventually the
wave packet spreads out over the classically allowed region rather uniformly with small speckles.
 This overall feature will no longer change, signaling that the system has dynamically equilibrated.

To illustrate this dynamical equilibration process, we compute how
the expectation of momentum changes with time for the ripple billiard system.
The results for both $p_x$ and $p_y$ are shown in Fig. \ref{momentum}, where  
the momenta are seen to relax to equilibrium values after a
short period of large fluctuations.

\begin{figure}
\includegraphics[width=0.49\textwidth ]{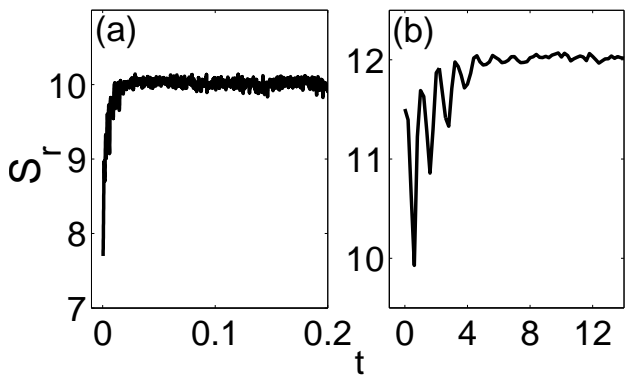}
\caption{Time evolutions of entropy in (a)  Henon-Heiles system;
(b) ripple billiard.}
\label{entropy}
\end{figure}

Equilibration should be accompanied by a maximizing entropy. Von Neumann was able to
define an entropy for a pure quantum state and proved that this entropy
will stay very close to its ensemble entropy almost all the time (the quantum H-theorem)~\cite{eV}. However, there appears no viable  procedure which one can use to compute this version of von Neumann entropy.  As an alternative, we define an entropy in the spirit of von Neumann,
\be
S_r=-\int \frac{|\psi(x,y,t)|^2}{|\psi_\infty(x,y)|^2}\ln \frac{|\psi(x,y,t)|^2}{|\psi_\infty(x,y)|^2} dxdy\,,
\ee
where $|\psi_\infty(x,y)|$ is the long time average of $|\psi(x,y,t)|^2$~\cite{zhuang_equilibration_2012}.
As shown in Fig.~\ref{entropy},  the entropy will increase  with time with small fluctuations and eventually saturate to a maximized value.  The increasing entropies in Fig.~\ref{entropy} can be regarded
as a ``spiritual" illustration of von Neumann's quantum H-theorem.

In summary, we have observed numerically that dynamical equilibration indeed occurs in
both the Henon-Heiles system and the ripple billiard: an initially localized Gaussian wave packet
with low entropy will dynamically evolve into a quantum state with a maximized
entropy, where the wave packet spreads out and looks irregular with speckles.
This is clearly consistent with both the quantum ergodic theorem and quantum H-theorem.
It is reasonable to expect that this kind of dynamical equilibration occurs in any quantum
chaotic system. Meanwhile we note that the equilibration process for the Henon-Heiles system
deserves more detailed study in the future. As noted before, the Henon-Heiles system
has no bound states. As a result, a Gaussian wave packet will eventually leak out 
and spread out in the whole space, not confined to just the triangle area, beyond the tunneling time.
In our numerical simulation, the equilibration time is clearly much shorter than the tunneling time.
It would be very interesting to investigate in which situations where the equilibration time becomes shorter
than the tunneling time. It might also be worthwhile  to formulate mathematically the ergordic theorem for this kind of systems.

\section{Verification of quantum ergodic theorem}
The mathematical expression of the quantum ergodic theorem is an inequality.
For an arbitrary operator $A$, this inequality reads~\cite{reimann_foundation_2008,short_equilibration_2011,short_quantum_2012}
\begin{equation}
\sigma^2_A\equiv \frac{\braket{| \mbox{tr}\{A\ket{\psi(t)}\bra{\psi(t)}\}- \mbox{tr}(A\rho_\infty)|^2}_t}{\|A\|^2}\le \frac{1}{d_{\mbox{eff}}}\,,
\label{ineq}
\end{equation}
where  $d_{\rm eff}\equiv1/\sum_k |a_k|^4$ measures effectively how many energy eigenstates
are occupied in the state $\ket{\psi}$.  The subscript $t$ in $\braket{}_t$ indicates a long time averaging. We emphasize that this inequality is much stronger than the following approximation
\be
\braket{\mbox{tr}\{A\ket{\psi(t)}\bra{\psi(t)}\}}_t\approx \mbox{tr}(A\rho_\infty)\,,
\label{ergodic}
\ee
which can be readily proved for any quantum systems with no energy degeneracy.  The above
approximation can still be true even when $\mbox{tr}\{A\ket{\psi(t)}\bra{\psi(t)}\}$ fluctuate
greatly from $\mbox{tr}(A\rho_\infty)$ as long as the positive large fluctuations cancel out
the negative large fluctuations. However, the fluctuations can not cancel each other in 
the inequality (\ref{ineq}); this means that the inequality (\ref{ineq}) dictates that the fluctuations are very small most of the time when $d_{\mbox{eff}}\gg 1$. So, the approximation relation (\ref{ergodic}) coupled with the inequality (\ref{ineq}) shows
that the long time averaging is equivalent to ensemble averaging, essence of
ergodicity, in all quantum systems that satisfy the condition (\ref{condition}).

\begin{figure}
\includegraphics[width=0.45\textwidth ]{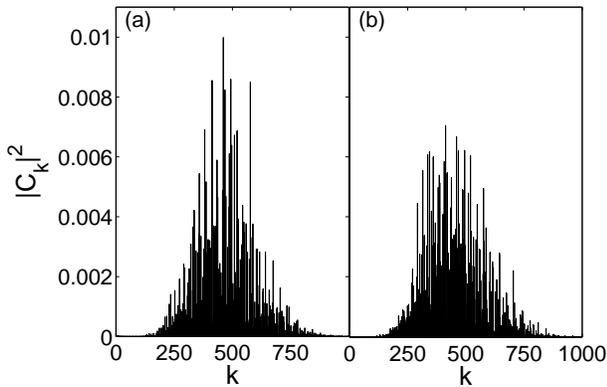}
\caption{The distribution of expansion coefficients $c_k$  in the ripple billiard system for (a) even-even eigenstates and (b) odd-even eigenstates.}
\label{decomposition}
\end{figure}

To test  numerically the inequality (\ref{ineq}), one needs to compute the energy eigenstates
$\ket{E_k}$ successively up to a high energy value and find the expansion coefficients
$c_k$. We are able to
do it for the ripple billiard system. The expansion coefficients of the initial state Eq. (\ref{initial_condition}) are computed and shown in Fig. \ref{decomposition}, where $c_k$'s are grouped 
according the symmetry of the eigenstates.  Both groups peak around 500th eigenstates and have a width
around 300. With the computed $c_k$'s, we find that the  effective dimension $d_{\rm eff}$
is around 300, satisfying the condition $d_{\rm eff}\gg 1$.

Without loss of generality, we choose to compute the left hand side (l.h.s.) of the 
inequality Eq. (\ref{ineq}) for momentum operator $\vec{P}=(p_x,p_y)$. By using the symmetry
of the eigenstates, one can readily show exactly that $\mbox{tr}(\rho_\infty p_x)=0$ and
$\mbox{tr}(\rho_\infty p_y)=0$. The time evolution of momentum is already shown in Fig.~\ref{momentum}. We calculate the long time average
between $t=10T_s$ and $t=14T_s$.  The time unit $T_s\equiv\frac{2(a+b)}{|p_i|/m}$ is the period of
the motion of a classical particle with the same initial momentum and position.
We find that $\braket{\mbox{tr}\{\ket{\psi(t)}\bra{\psi(t)}p_x\}}_t\simeq0.0021$ and $\braket{\mbox{tr}\{\ket{\psi(t)}\bra{\psi(t)}p_y\}}_t\simeq-0.0033$, very close to the ensemble average $\mbox{tr}(\rho_\infty p_x)=0$ and  $\mbox{tr}(\rho_\infty p_y)=0$, respectively.  At the same time,  we find that
\ba
&&\braket{| \mbox{tr}\{\vec{P}\ket{\psi(t)}\bra{\psi(t)}\}- \mbox{tr}(\vec{P}\rho_\infty)|^2}_t\nonumber\\
&&=\braket{| \mbox{tr}\{\vec{P}\ket{\psi(t)}\bra{\psi(t)}|^2}_t \simeq0.0039\,.
\ea

\begin{figure*}
\centering
\subfigure[]{
\label{correspondence}
\includegraphics[width=0.45\textwidth]{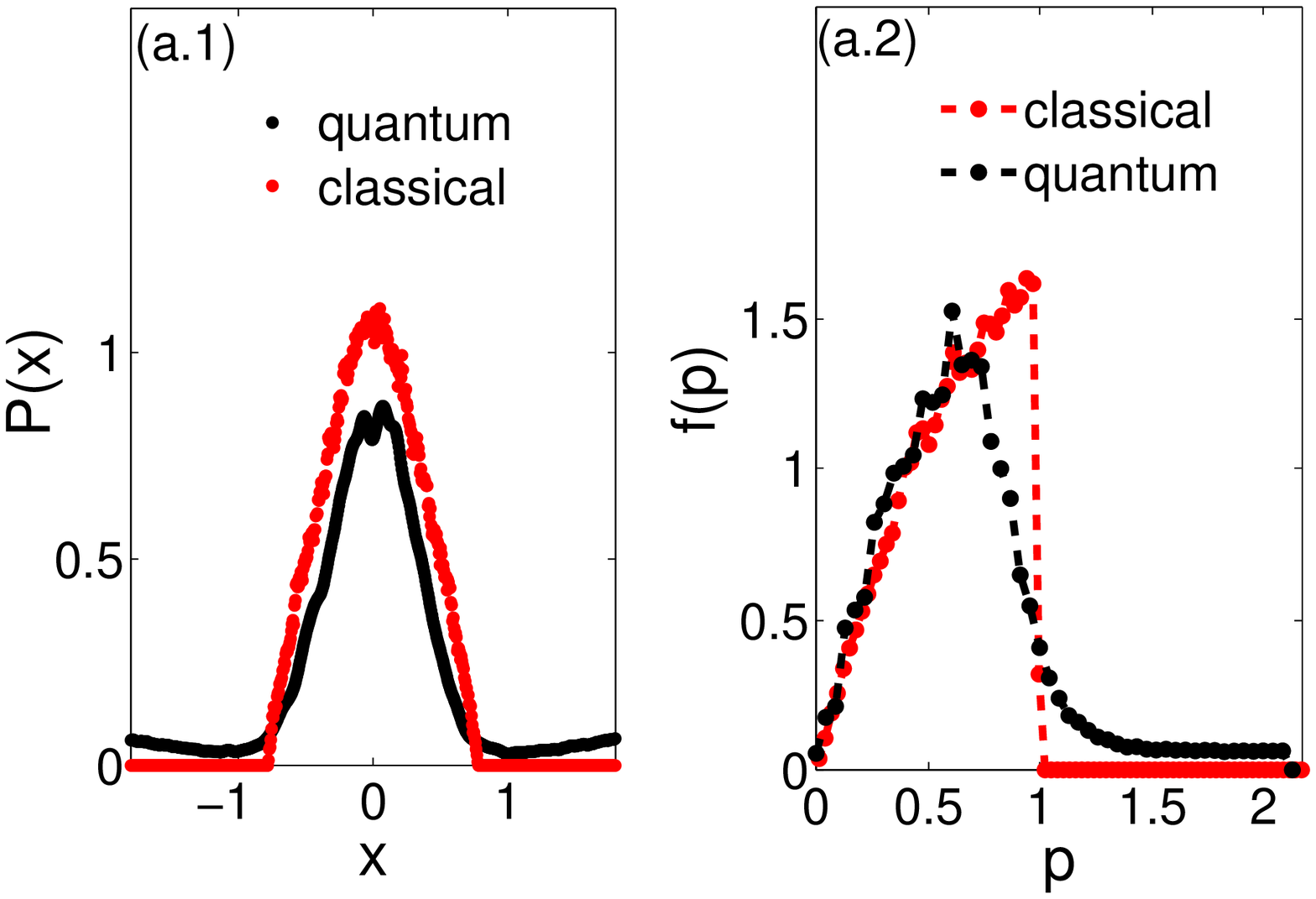}}
\centering
\subfigure[]{
\label{husimiplot}
\includegraphics[width=0.4\textwidth ]{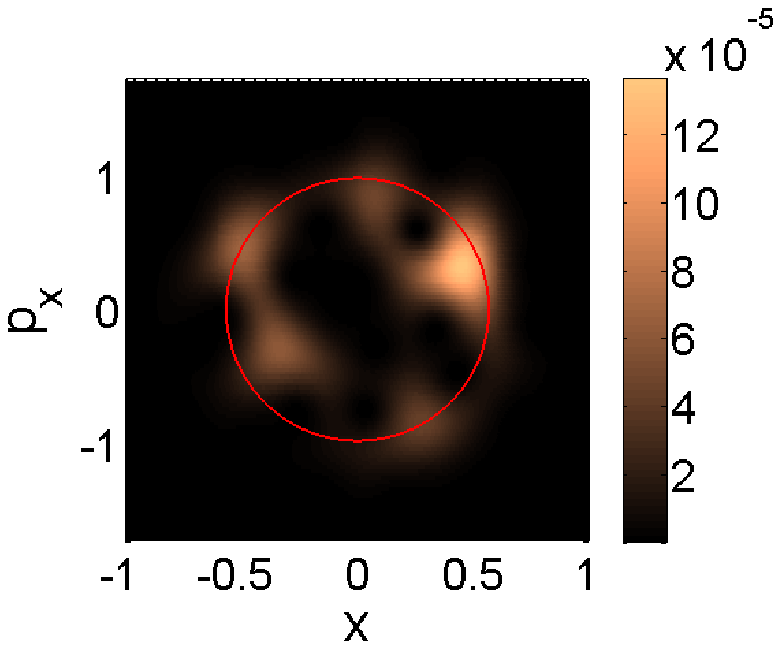}}
\caption{
(a) Comparison between   quantum and classical  equilibrium distributions
in the Henon-Heiles system.  $P(x)$ is for the real space and $f(p)$ for momentum space.
The unit for $x$ is $r_c$ and the unit for $p$ is $p_0$.
(b) Section of the Husimi function $H_{r.p}\equiv \braket{\vec{r},\vec{p}|\hat{\rho}(t) |\vec{r},\vec{p}}$ in the Henon-Heiles system at $t=0.2126$. The red circle is the section of the classical phase space density $\rho_c(\vec{r},\vec{p})$. The section is at $y=0, p_y=0$. The unit for $x$ is $r_c$ and the unit for $p_x$ is $p_0$.  The coarse-graining parameter for the Husimi function $\sigma\simeq0.11r_c$;
}
\end{figure*}

For operator $\vec{P}$, the maximum value $\|\vec{P}\|^2=\sup\{\braket{\psi|P^\dagger P|\psi}\}$.
Since we have chosen $m=1/2$ and the wave function is only none zero inside the billiard, we have
 $\|\vec{P}\|^2=\sup\{\braket{\psi|H|\psi}\}$, that is, $\|\vec{P}\|^2$ is effectively
 the largest energy in the occupied Hilbert space. According to Fig. \ref{decomposition}, the ocuppied Hilbert space is roughly spanned by the eigenstates  between the100th even-even(odd-even) eigenstates and the 1000th even-even(odd-even) eigenstates. The upper bound can be estimated as the eigenvalue of the 1000th even-even(odd-even) eigenstates, i.e. $\|\vec{P}\|^2\simeq 55.64$. So, the relative fluctuation or the l.h.s. of Eq. (\ref{ineq}) is $\sigma_{\vec{P}}^2\simeq 7\times 10^{-5}$.  As $1/d_{\mbox{eff}}\simeq 3\times10^{-3}$,  we see that
the inequality (\ref{ineq}) is clearly satisfied. Moreover, our computation in fact shows
that the right hand side (r.h.s.) of the inequality is  about 30 times larger than the l.h.s. This indicates that it is possible to improve the inequality, for example, replacing $\|A\|^2$ with the averaged value.

Before we proceed further,  we want to mention that the inequality for quantum ergodicity
proved by von Neumann ~\cite{eV} has a different
upper bound on the r.h.s.. However,  his theoretical formalism relies
on the introduction of macroscopic operators (such as macroscopic position and momentum) that commute with each other.  It appears not be a straightforward task to construct these macroscopic operators and compute them numerically. As a result, we did
not compute von Neumann's upper bound.

\section{Quantum-classical correspondence at equilibrium}
According to the quantum ergodic theorem,  the equilibrium state is described by
the density matrix $\rho_\infty$, which can be regarded as the quantum  micro-canonical ensemble.
We analyze  this quantum equilibrium state
and find that it possesses many features that resemble the classical  micro-canonical ensemble.
This kind of quantum-classical correspondence has been studied and found in a spin
system~\cite{Emerson,Guarneri}.  To illustrate  this  quantum-classical correspondence, we compare $\rho_\infty$ in both the real space and the momentum space with the classical micro-canonical ensemble
\be
\rho_c= \frac{1}{\Omega}\delta(H(\vec{p},\vec{r})-E)
\ee
where $\Omega$ is the normalization factor and $H(\vec{p},\vec{r})$ is the corresponding classical Hamiltonian of the fully chaotic system. Here we choose the Henon-Heiles system to analyze because its probability distribution in phase space is more general than the billiard system where $|\vec{p}|$ is a constant of motion. For the quantum results, we calculate $n_\infty(\vec{r})=\braket{\vec{r}|\rho_\infty|\vec{r}}$ and $n_\infty(\vec{p})=\braket{\vec{p}|\rho_\infty|\vec{p}}$ by long-time averaging; For the classical results, we calculate the probability distribution in real space and momentum space $n_c(\vec{r})$ and $n_c(\vec{p})$ from the micro-canonical ensemble $\rho_c(r,p)$ by integration over $\vec{p}$ and $\vec{r}$ separately.

For better comparison, we choose the following marginal distribution without loss of generality: for the real space,
we integrate out  the $y$ dimension to obtain the density distribution $P(x)$;
for the momentum space,  we integrate out  the angle variable to find the momentum distribution $f(p)$. The  results are shown in Fig. \ref{correspondence}, where we see the quantum and classical distributions are consistent, except for some quantum tunneling effect indicated by the non-zero value of the quantum distribution in the classically forbidden region.

This correspondence also exists in phase space. In quantum mechanics, the
uncertainty relation does not allow the construction of a phase space in principle.
However, a kind of quasi-quantum distribution  in phase space can be constructed
with the  Husimi function $H_{r.p}$ ~\cite{lee_signatures_1993,toscano_husimiwigner_2008}.
This is to calculate the projection of the density operator on a Gaussian wave packet $\braket{\vec{r^\prime}|\vec{r}, \vec{p}}=C\exp(-\frac{(\vec{r^\prime}-\vec{r})^2}{2\sigma^2}+\frac{i\vec{p}\cdot(\vec{r^\prime}-\vec{r})}{\hbar})$ centered around position $\vec{r}$ and momentum $\vec{p}$
\be
H_{r.p}\equiv \braket{\vec{r},\vec{p}|\rho|\vec{r},\vec{p}}\,.
\ee
The  Husimi function can be understood as coarse grained phase space density with parameter $\sigma$ controlling the coarse graining.  We have computed the Husimi function $H_{r,p}$ for
a density matrix $\rho\equiv \ket{\psi}\bra{\psi}$ at $t=0.2126$ (which is after equilibration) and compared
it to the classical ensemble $\rho_c$. For easy illustration, we use a 2-dimensional section
in the 4-dimensional phase space of the Henon-Heiles system. Without loss of generality, we choose the section at $y=0, p_y=0$ and the results are
plotted in Fig. \ref{husimiplot}.  We can see that the Husimi function centers around the phase space where the classical density is non-zero, indicating the quantum-classical correspondence.  Note
that the quantum fluctuations in the phase space are much larger than the ones seen in Fig. \ref{correspondence}. The reason is that the results in Fig. \ref{correspondence} are obtained after being
averaged over time and integrated over a given dimension.

One possible  understanding for this correspondence is through the quantum and classical Liouville equations. Intuitively, one can think of the quantum wave-packet as an ensemble of particles with equivalent classical probability density in the phase space~\cite{ballentine_inadequacy_1994,ballentine_moment_1998}.  The time evolution of quantum density matrix operator $\rho(t)$ is governed by the quantum Liouville equation,  $\partial_t\rho(t)=\frac{1}{i\hbar}[{H},{\rho(t)}]$; the time evolution of the classical probability density $\rho_c(t)$ is governed by the classical Liouville equation,  $\partial_t\rho_c(t)=[H,\rho_c(t)]_{\rm PB}$. These two time evolution equations have an identical algebraic structure, implying a possible quantum-classical correspondence in the equilibrium states. Moreover, there are numerical evidences for this correspondence in the studying of
the time evolution of the ensemble average and quantum expectation value~\cite{ballentine_inadequacy_1994,ballentine_moment_1998}. Note that
this quantum-classical correspondence is not implied in  the quantum ergodic theorem~\cite{eV,short_equilibration_2011,short_quantum_2012,reimann_foundation_2008,reimann_equilibration_2012}. 

\begin{figure}
\centering
\subfigure[]{
\label{finalstate} 
\includegraphics[width=0.22\textwidth ]{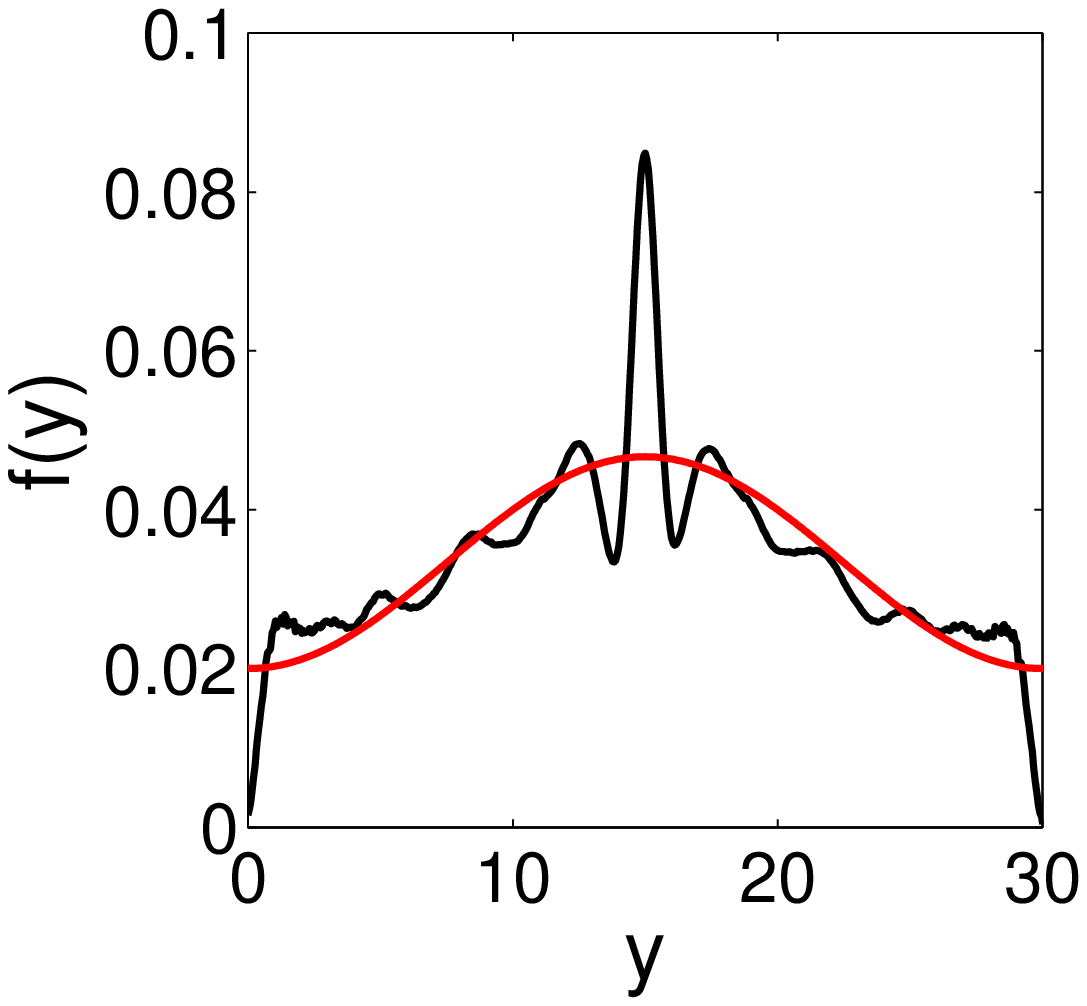}}
\centering
\subfigure[]{
\label{finalstaten} 
\includegraphics[width=0.24\textwidth ]{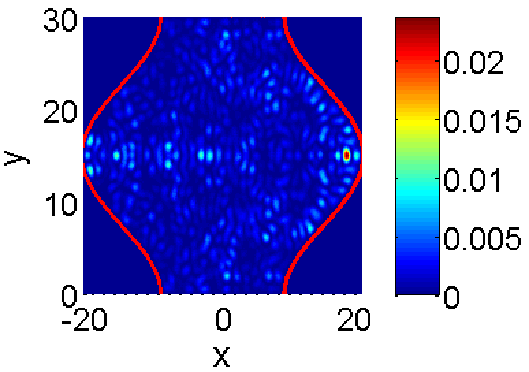}}
\centering\\
\subfigure[]{
\label{eigenfy} 
\includegraphics[width=0.22\textwidth ]{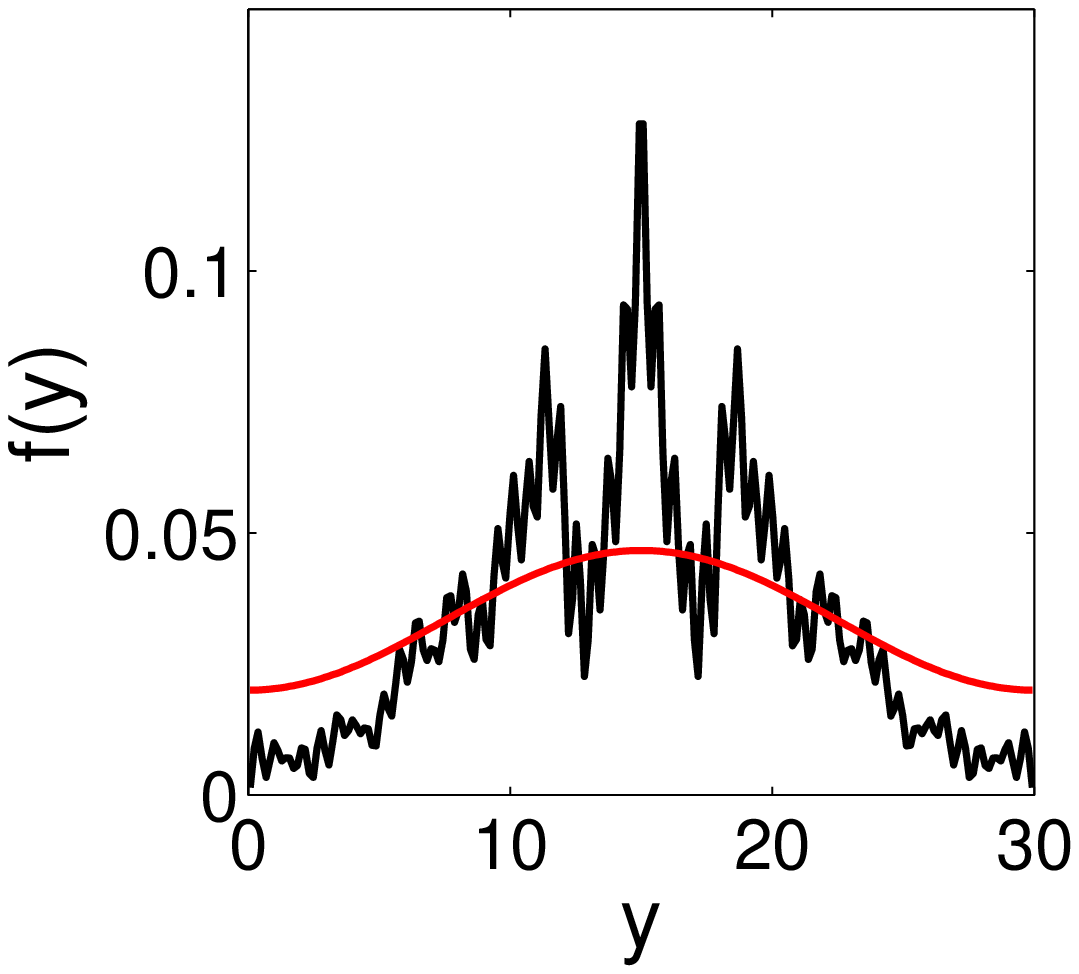}}
\centering
\subfigure[]{
\label{eigen} 
\includegraphics[width=0.24\textwidth ]{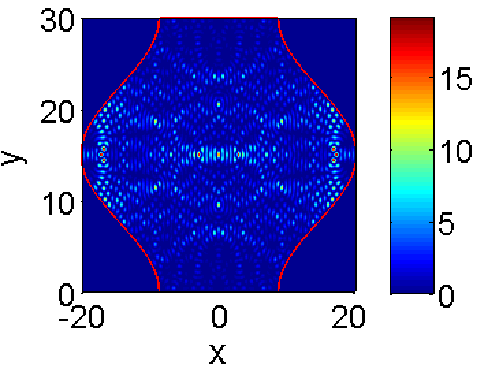}}
\caption{(a) Marginal distributions $f(y)\int n(\vec{r}) dx$ for the equilibrium state  in the ripple billiard system. The black line is for the quantum result while the red line is for the classical result.
(b) The density distribution $n(r)$ of the equilibration state in the ripple billiard system.
(c) Marginal distribution $f(y)$ for the 493rd even-even eigenstate  in ripple billiard system.
(d) Density distribution $n(x,y)$ of the 493rd even-even eigenstate. For this eigenstate,  the expansion  coefficient $|c_i|^2=0.0086$ in the initial Gaussian wave packet (\ref{initial_condition}).
}
\end{figure}

We emphasize here that the quantum-classical correspondence discussed so far needs to be
understood in the sense of typicality~\cite{Lebowitz_pt,popescu_entanglement_2006,popescu_foundations_2005,
goldstein_canonical_2006,reimann_typicality_2007}. As dictated by the quantum ergodic theorem,
the equilibrium state is completely determined by the initial expansion coefficient $|c_k|^2$.
For a typical initial state, $|c_k|^2$ should have a distribution similar to what is shown in
Fig. \ref{decomposition} with a large effective dimension $d_{\rm eff}$. For these states,
we expect that the quantum-classical correspondence hold. However, for many atypical states,
this quantum-classical correspondence may not hold. For example, the ergodic inequality (\ref{ineq})
holds for an energy eigenstate; but energy eigenstate does not belong to the typicality class.
Here we use the ripple billiard to illustrate this point as the energy eigenstates in this system
can be computed successively~\cite{li_quantum_2002,xiong_universal_2011}.
We study the position space marginal distribution $f(y)=\int \braket{\vec{r}|\rho|\vec{r}} dx$, and compare it with the classical result $(b-a\cos(2\pi y/L))/bL$. We compute $f(y)$ first for the equilibrium state $\rho(t\gg T_s)$. We see in Fig. \ref{finalstate} that  the quantum result (black line) are in good agreement with the classical result (red line).   For completeness,  we also show the density $\braket{\vec{r}|\rho|\vec{r}}$ in Fig. \ref{finalstaten}.

For eigenstates, we choose to study the eigenstates that have relatively large expansion coefficients
$|c_k|^2$ in the initial wave packet (\ref{initial_condition}). We find that the results
for them are similar. Without loss of generality, we choose
the 493rd even-even eigenstates with  coefficient $|c_k|^2=0.0086$ as the example.
The marginal distribution $f(y)$ for this eigenstate is shown in Fig. \ref{eigenfy}. Compared with the result for the equilibrium state in Fig. \ref{finalstate}, we see a clear distinction: the eigenstate has larger fluctuations and bigger deviation from the classical distribution. In fact,  this
distinction is also seen in the density plot.  As seen in Fig. \ref{eigen}, the density distribution
for the eigenstate has certain uneven patterns, similar to  what have been found in ``scar"
states~\cite{heller_bound-state_1984}; these patterns are absent in  the distribution of the equilibrium state ( Fig. \ref{finalstaten}). Although eigenstates appear to be too special, their difference
from a typical quantum state  does indicate that  not all quantum states that
satisfy the inequality (\ref{ineq}) can relax to an equilibrium state which has the quantum-classical correspondence. In other words, all quantum states in a chaotic system may be classified into
two groups: one has the quantum-classical correspondence and the other does not have. It is not
yet clear whether there are simple ways to separate them other than using direct computation as we did. In Ref.\cite{Guarneri},  a remarkable correspondence between eigenstates and their
classical counterparts is found for the shape of eigenfunctions and
 the local spectral density of states. This seems to indicates that when the
 properties not related to the complexity of a wave function is considered,  
 the quantum-classical correspondence can still be found for eigenstates.

\section{Fluctuation property}
\begin{figure}
\centering
\subfigure[]{
\label{fig:subfig:b} 
\includegraphics[width=0.23\textwidth ]{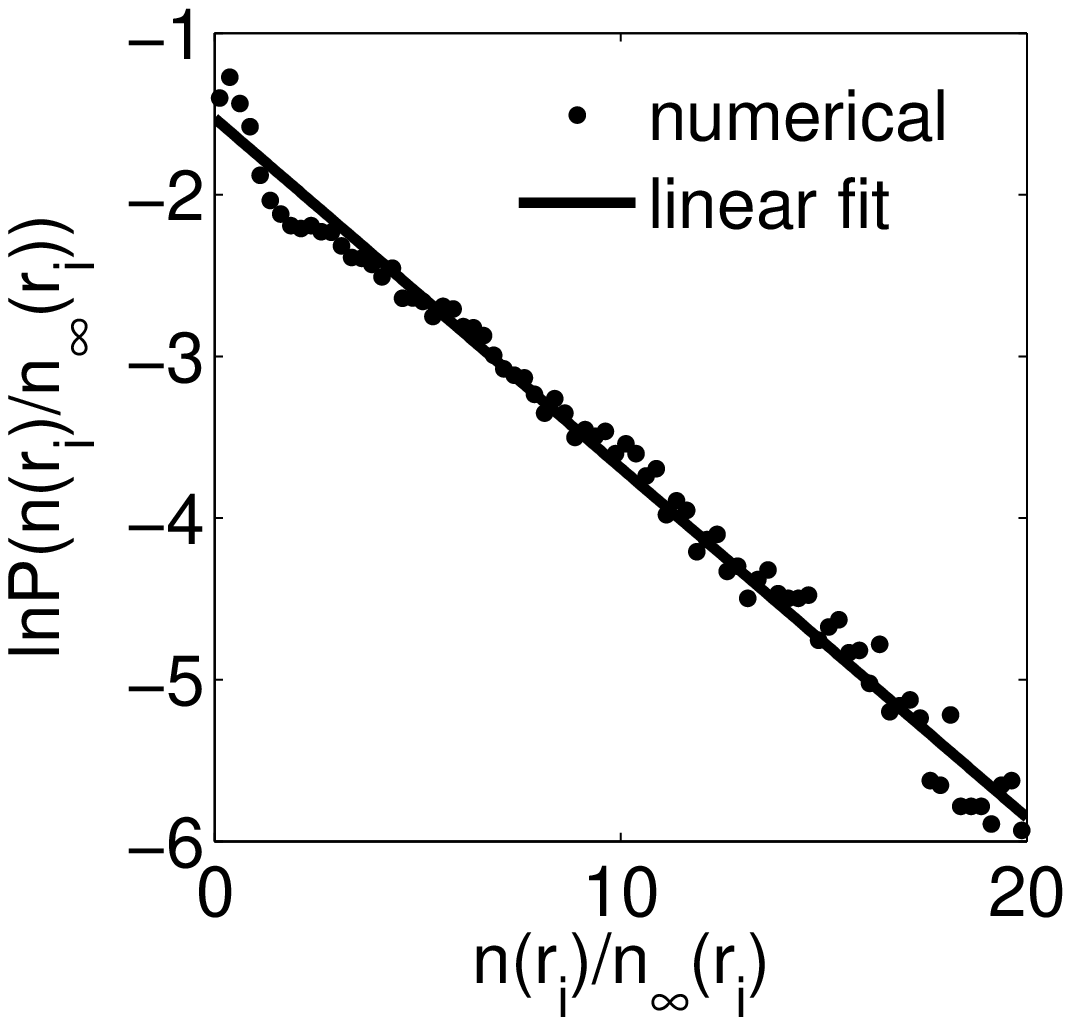}}
\centering
\subfigure[]{
\label{fig:subfig:b} 
\includegraphics[width=0.23\textwidth]{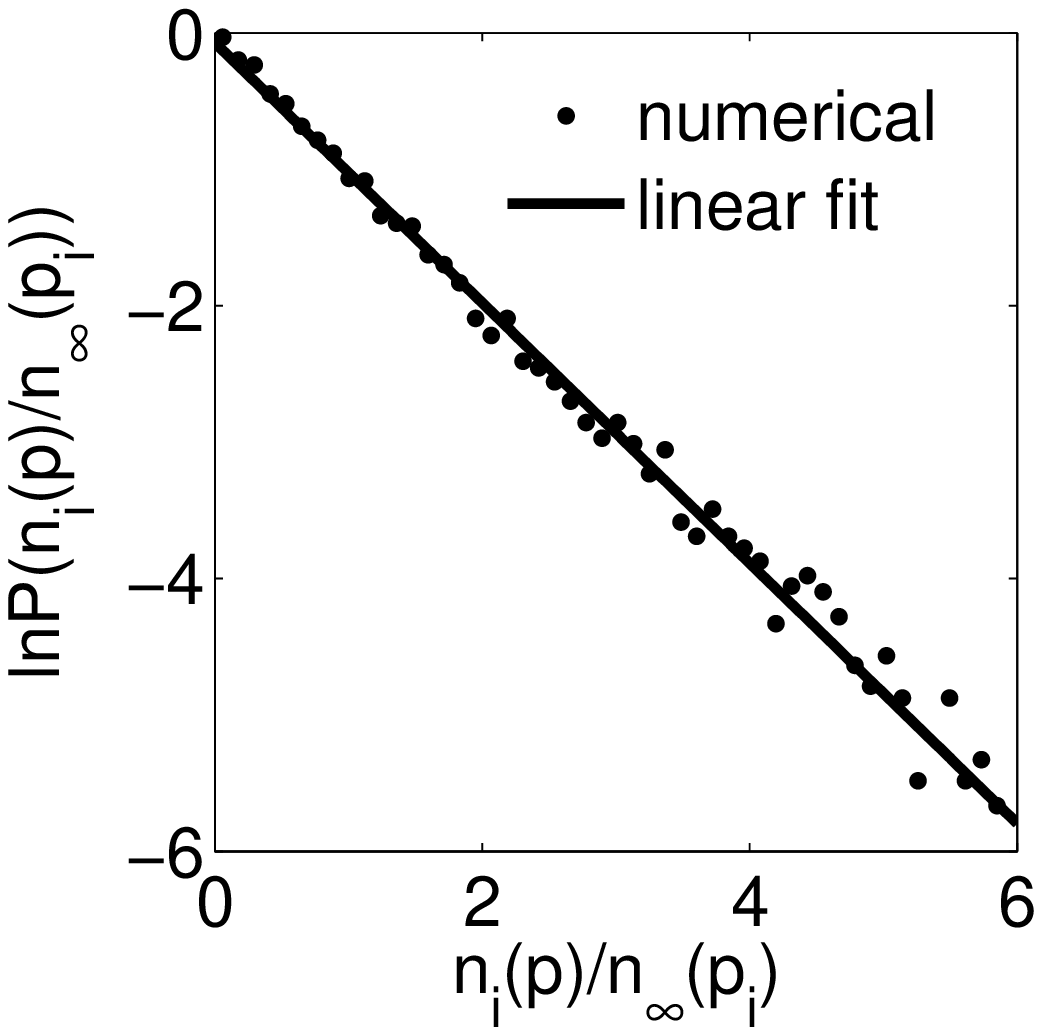}}
\centering \\
\subfigure[]{
\label{fig:subfig:b} 
\includegraphics[width=0.23\textwidth ]{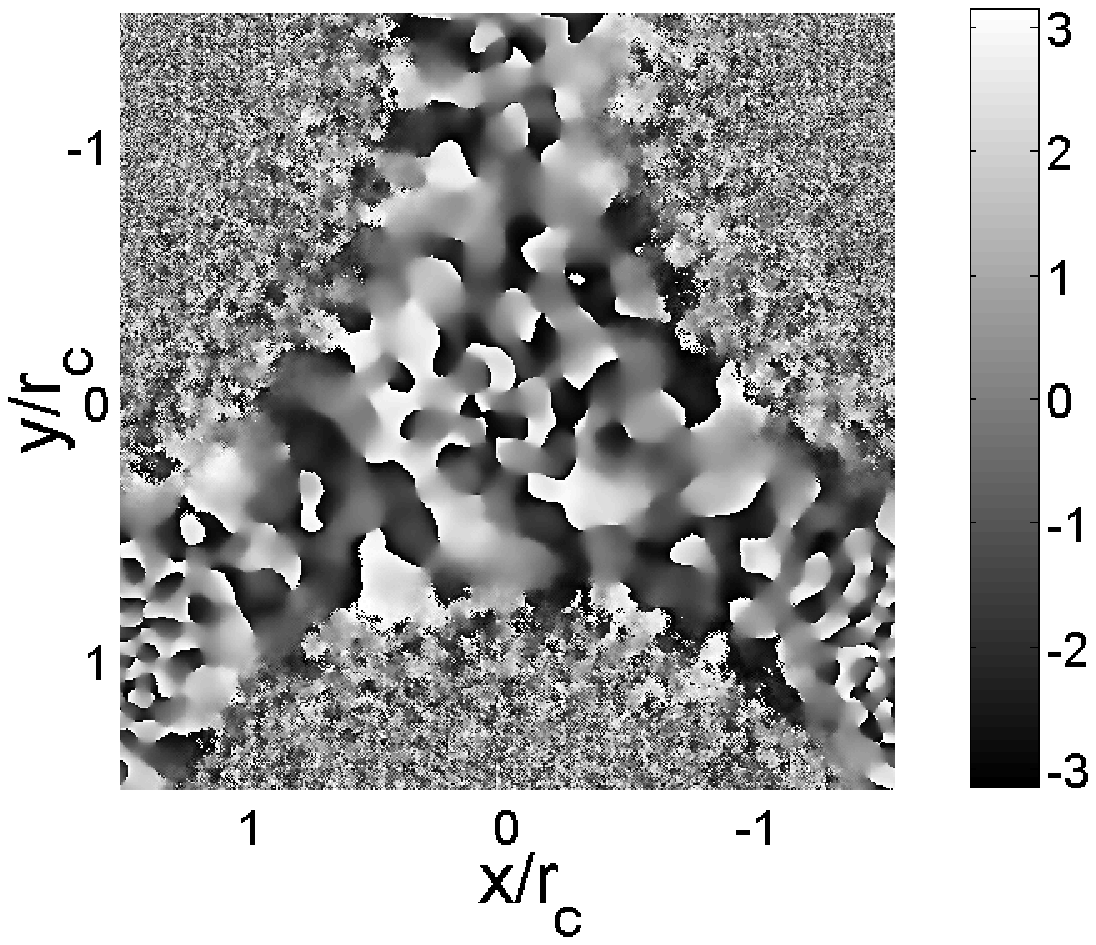}}
\centering
\subfigure[]{
\label{fig:subfig:b} 
\includegraphics[width=0.23\textwidth]{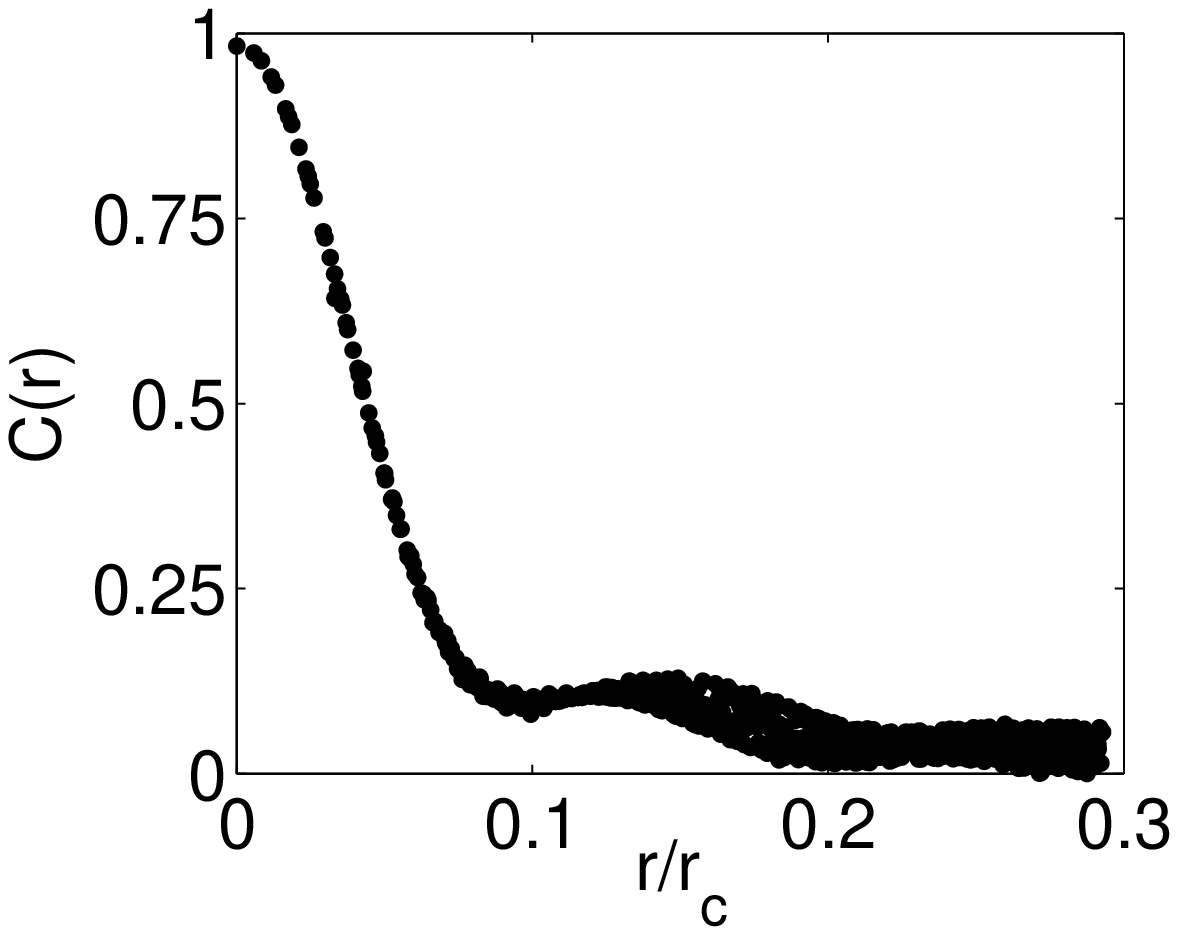}}
\caption{
(a) Log plot of the distribution function $P(n(r_i)/n_\infty(r_i))$ of the relative {\it quantum} fluctuation $n(r_i)/n_\infty(r_i)$.
(b)log plot of the distribution function $P(n(p_i)/n_\infty(p_i))$ of the relative {\it quantum} fluctuation $n(p_i)/n_\infty(p_i)$.
(c) The phase of the equilibrated wave function in the real space for the Henon-Heiles system.
(d) The normalized spatial correlation function $C(r)=\frac{1}{\braket{\Delta n^2}}\braket{(n(0)-\bar{n})(n(r)-\bar{n})}$ of density in the real space at equilibrium for the Henon-Heiles system.
}
\label{exponential}
\end{figure}

While the equilibrium density operator $\rho_\infty$ shares a good correspondence with the classical microcanonical ensemble $\rho_c$, the operator $\rho(t)$ does fluctuate around the equilibrium density operator, as shown in Fig. \ref{finalstaten}. This fluctuation has been shown to obey the exponential distribution in the ripple billiard system~\cite{xiong_universal_2011}. Here we focus on the Henon-Heiles system to bring this exponential distribution beyond billiard systems. To quantitatively show the fluctuation, we compare the probability density in both the real space and the momentum space, i.e. we compare $n(\xi)\equiv \braket{\xi|\rho(t)|\xi}$ with $n_\infty(\xi)$, $\xi=\vec{r},\vec{p}$.

Numerically this is realized by calculating the density $n(\xi_i) \mbox{ and }n_\infty(\xi_i) (i=1,\cdots, N)$ of $N$ small discrete region and calculate the distribution of the relative fluctuation $n(\xi_i)/n_\infty(\xi_i)$. Due to limited numerical precision, we choose different $N$ in the real space and momentum space and the results are shown in Fig. \ref{exponential}(a,b). Both distributions are found to be well fitted by exponential distribution. Due to the different $N$, the coefficients of the fitted exponential distributions are different in the real space and the momentum space.

\begin{figure}
\centering
\subfigure[]{
\label{fig:subfig:b} 
\includegraphics[width=0.23\textwidth]{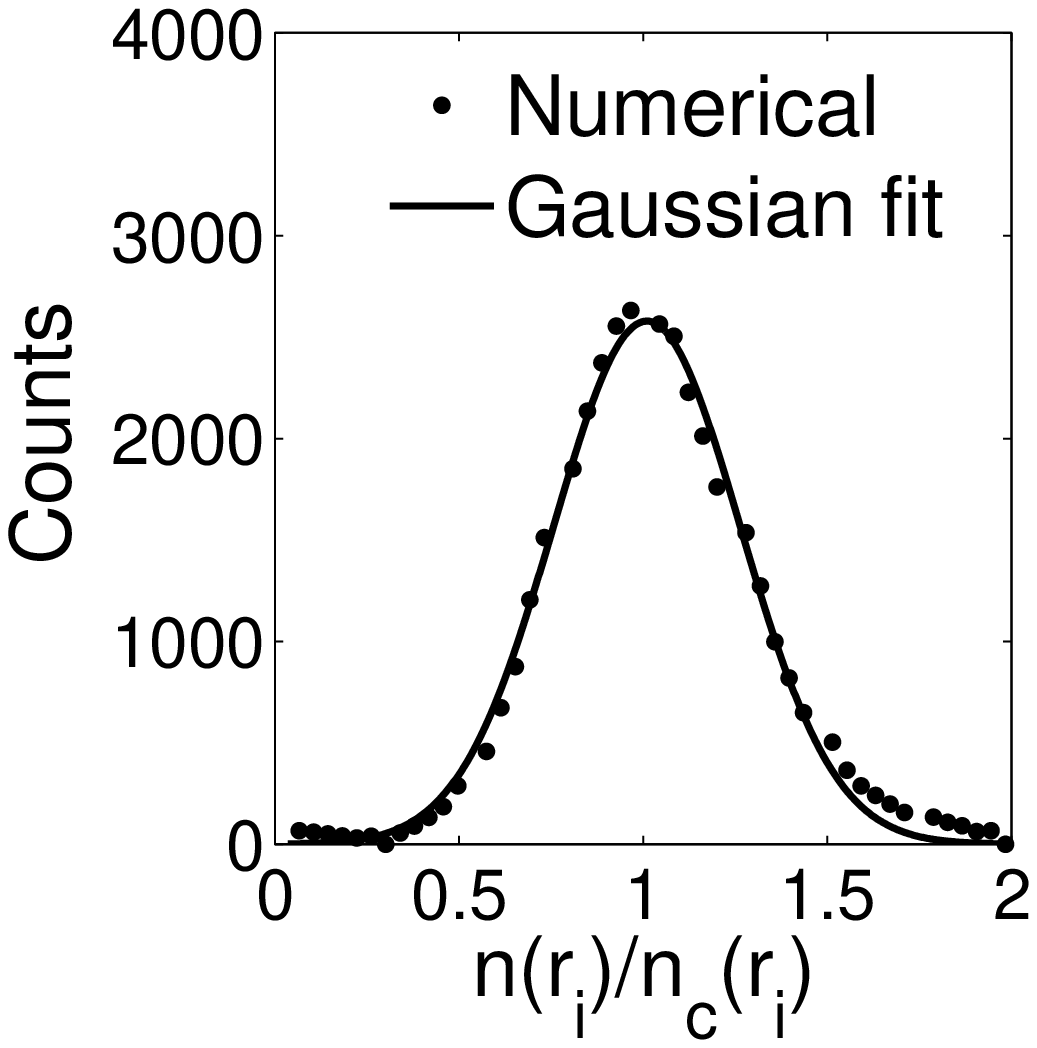}}
\centering
\subfigure[]{
\label{fig:subfig:b} 
\includegraphics[width=0.23\textwidth]{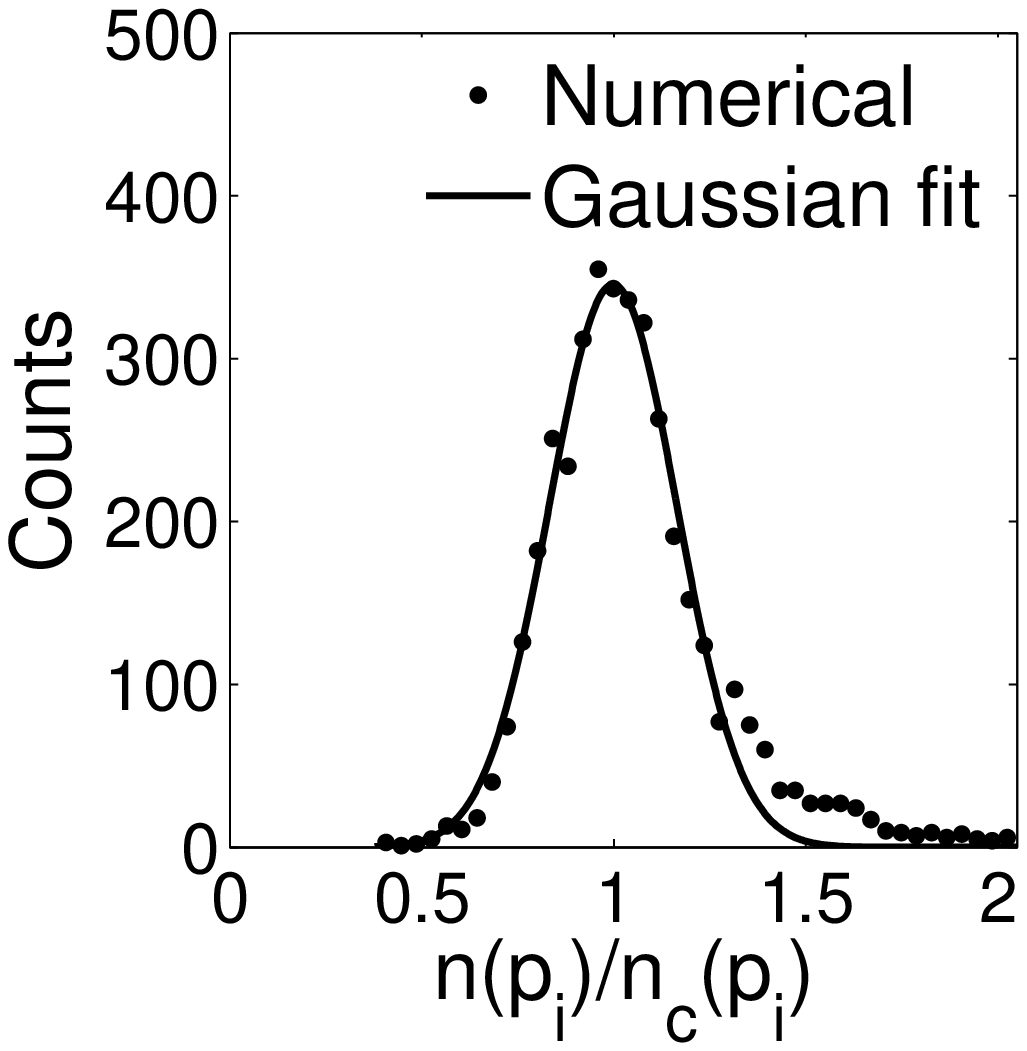}}
\caption{
(a) The distribution function $P(n(r_i)/n_\infty(r_i))$ of the relative {\it classical} fluctuation $n(r_i)/n_\infty(r_i)$.
(b) The distribution function $P(n(p_i)/n_\infty(p_i))$ of the relative {\it  classical} fluctuation $n(p_i)/n_\infty(p_i)$.
}
\label{classical}
\end{figure}

This exponential distribution results from the randomness of wave function in a quantum chaotic system. Though the system is deterministic and evolves strictly with the Schr\"{o}dinger equation, the  scattering by 
the potential wall bring the wave function to a state where the probability density and phases behave like random numbers.  Here we demonstrate the randomness by computing the phase of the wave function and the normalized probability density  correlation function $C(r)=\frac{1}{\braket{\Delta n^2}}\braket{(n(0)-\bar{n})(n(r)-\bar{n})}$ in the real space after equilibrium in the Henon-Heiles system. The results are shown in Fig. \ref{exponential}(c, d). We see that the phases vary rapidly in real space and the spatial correlation in real space decreases exponentially to zero within small distance compared to the
characteristic length scale $r_c$. These features  indicate certain randomness in the behavior of wave function in quantum chaotic systems. With this established randomness, the proof in Ref.~\cite{xiong_universal_2011} with slight modification, can explain this exponential distribution (see also the Appendix). Note that this randomness of wave function in quantum chaotic system is consistent with the idea of typicality~\cite{goldstein_canonical_2006}. It also supports, though not directly, the assumption we made in Eq. (2) and Eq. (3) in our previously work~\cite{zhuang_equilibration_2012}. 

This exponential fluctuation is a pure quantum phenomenon. To show the quantum nature we consider the classical correspondence of an ensemble of particles obeying the same initial Gaussian
distribution of position and momentum and calculate their time evolution. Their phase space  distribution $\rho_c(t)$ approaches the microcanonical distribution $\rho_c$ due to the ergodicity in the Henon-Heiles system with certain fluctuation. Following the quantum case, we calculate the relative fluctuations $n_c(\xi,t)/n_c(\xi), \xi=\vec{r},\vec{p}$ and study their distributions. The classical relative fluctuations is Gaussian  as indicated in Fig. \ref{classical}. This shows that the exponential distribution of relative fluctuations is of quantum nature and absent in the classical system.

Besides the difference from classical case, we note that this exponential distribution is also different from the Porter-Thomas distribution in eigenstates~\cite{porter_fluctuations_1956} and the result now known
as the Berry's conjecture~\cite{Berry}. These results are only for eigenstates, which are real up to
an overall phase for systems with time reversal symmetry.  
 The exponential distribution only exists for a complex wave function, For a generic quantum system,
 even if its all energy-eigenstates are real, its dynamical wave function is in general complex.
This distinction again shows that the general equilibrium statistical properties cannot be understood simply by properties of single eigenstates.   Superposition of numerous eigenstates is crucial.
We feel it very helpful to compare directly 
the exponential distribution and the the Port-Thomas distribution~\cite{porter_fluctuations_1956}. We have re-derived these two  distributions 
within the same mathematical framework in the Appendix.

\section{Discussion and Summary}
In summary, we have studied equilibration of quantum chaotic systems in the framework
suggested by von Neumann in 1929. Our study has examined various aspects of this issue,
such as dynamical equilibration,  quantum-classical correspondence between quantum equilibrium density operator $\rho_\infty$ and classical microcanonical ensemble $\rho_c$,  and the relative fluctuations around the equilibrium state.  All the results are illustrated with two single particle
quantum chaotic systems. We  expect most of the results  to hold in many-body quantum chaotic systems.

The dynamical equilibration should hold in many-body quantum chaotic systems because
it is completely determined by the energy gap statistics, which is universal
for all quantum chaotic systems. The quantum-classical correspondence should also hold
as it apparently originates from the similarity between quantum and classical Liouville equations.

However, the fluctuation properties in a many-body system will generally have Gaussian form, different from the single-particle case. The only except might be a Bose-Einstein condensate or a superconductor
where the many-body system can be described by a single-variable wave function. In all other cases,
the probability density at a given point $\xi=\vec{r},\vec{p}$ can be expressed by
\be
n(\xi_1)=\int d\xi_2\cdots d\xi_N\, n(\xi_1,\xi_2\cdots,\xi_N)\,,
\ee
where $N$ is the number of particles in this system. The fluctuation in $n(\xi_1,\xi_2\cdots,\xi_N)$ should have the exponential distribution due to the quantum chaotic nature of the many-body system. The integration over $N-1$ variables will produce a Gaussian distribution
due to the central limit theorem.

We note that one has to use many-body systems to discuss
the canonical statistics for a subsystem of the large isolated system. However, one
can use  the correspondence between $\rho_\infty$ and $\rho_c$ illustrated here with single-particle
systems to show that a subsystem of the many-body system will behave in canonical ensemble statistics. The derivation is similar to the work by Srednicki~\cite{srednicki_chaos_1994} and here we do not discuss further.

Although von Neumann has laid down the basic theoretical framework for the dynamical
equilibration in quantum systems in 1929~\cite{eV}, we believe that the study of this
fundamental issue has just started and many basic questions are still await to be answered.
Our study here and other related studies~\cite{Santos,Santos2} have indicated that
the quantum ergodic theorem hold for a class of quantum systems larger than
what is specified by the condition (\ref{condition}).  It is imperative to know how
to relax the condition (\ref{condition}) mathematically. We are also not clear about
the properties of the entropy  defined for a pure state by von Neumann.
We believe that these studies will not only lead to a better perspective on the foundation of quantum statistical
mechanics but also produce new physics such as the quantum multi-temperature equilibrium
state predicted in Ref.~\cite{zhuang_equilibration_2012} by us.

\begin{acknowledgments}
This work is supported by the NBRP of China (2012CB921300,2013CB921900) and
the NSF of China (11274024,11128407) and the RFDP of China (20110001110091).
\end{acknowledgments}

\appendix*
\section{Derivation of Exponential Distribution and Port-Thomas Distribution}
Here we give detailed derivations for  the exponential distribution of equilibrium wave function~\cite{xiong_universal_2011} and the Port-Thomas distribution of eigenstate wave function~\cite{porter_fluctuations_1956} with the assumption that wave-function in quantum chaotic systems achieve the maximum randomness subject to the restriction of normalization and other symmetry requirements. This assumption follows naturally from our numerical observations(Fig.~\ref{exponential} (c), (d)) and the spirit of random matrix theory, and was used by Berry 
in his study of eigen-wave functions in chaotic systems~\cite{Berry}
It is important to  note the difference between equilibrium wave functions and 
eigenstates wave functions is that equilibrium wave functions are generally complex 
and eigenstates wave functions are real up to an overall phase for systems with time reversal symmetry. This key difference leads to the different distributions. For simplicity we restrict our derivation for real space wave functions in chaotic billiards. Extension to other systems and momentum space can be achieved (however, without strict mathematical rigorousness) by renormalizing densities with averaged values.
 
For a  wave function $\psi$, we discretize the system area into $N$ equal infinitesimal pieces with their centers located at $\vec{r_i}, i=1,\cdots,N$ and denote $\alpha_i=\psi(\vec{r_i})\sqrt{\frac{A}{N}}$, where $A$ is the total area of the system. It follows from normalization that $\sum_{i=1}^N |\alpha_i|^2=1$. The notation here is the same with Ref.\cite{xiong_universal_2011}. For a complex wave function in dynamical evolution $\alpha_i$ is complex; for an eigenfunction $\alpha_i$ is real. In the following  we derive the distribution of $|\alpha_i|^2$ for both cases. 

Note that the following proof is given without consideration of the discrete symmetries:  (1) reflection symmetry initial state(Eqn. \ref{initial_condition}) in ripple billiard system; (2) reflection symmetries in eigenstates of ripple billiard  system, and (3) the reflection symmetries in eigenstates of Henon-Heiles system. This is because these symmetries can be treated by simply assuming full randomness of a wave function in a smaller area while changing the normalization, respectively, by $1/2$(Eq. \ref{initial_condition})), $1/4$ (ripple billiard), or $1/6$(Henon-Heiles).  

\subsection{Proof of exponential distribution of equilibrium wave function~\cite{xiong_universal_2011}}
Consider $N$ complex numbers $\alpha_i$ that satisfies normalization condition $\sum_{i=1}^N |\alpha_i|^2=1$ and each complex number is equivalent to another. Suppose the $N$ complex numbers are fully random subject to the normalization condition. This implies that each state 
$\{\alpha_1,\alpha_2,\cdots,\alpha_N\}$ is of equal possibility on the hypersphere $\sum_{i=1}^N|\alpha_i|^2=1$. So the distribution of the amplitude  of one complex number $\alpha_j$ at 
$|\alpha_j|^2=\gamma$ is

\be
P(\gamma)=\frac{\int d^2\alpha_1 \cdots d^2\alpha_N\delta(\gamma-|\alpha_j|^2)\delta(1-\sum_{i=1}^N|\alpha_i|^2)}{d^2\alpha_1 \cdots d^2\alpha_N \delta(1-\sum_{i=1}^N|\alpha_i|^2)}
\ee

Let  $x=|\alpha_j|^2$. The denominator equals the surface area of $2N$-dimensional hypersphere. The numerator includes a factor of a $(2N-2)$-dimensional hypersphere with radius $\sqrt{1-x}$. Recall the $2N$-dimensional hypersphere of radius $R$ has an area of $S_{2N}(R)=\frac{2\pi^N}{\Gamma(N)}\times(R)^{2N-1}$.  We have 
\ba
P(\gamma)&=&\frac{\int_0^\infty \pi dx\delta(x-\gamma)\frac{2\pi^{N-1}}{\Gamma(N-1)}(1-x)^{\frac{2N-3}{2}}}{2\pi^N/\Gamma(N)}\nonumber\\
&=&(N-1)(1-\gamma)^{N-3/2}\,.
\ea
Let $\gamma=\frac{A}{N}n(\vec{r})$, where $n(\vec{r})=|\psi(\vec{r})|^2$ is 
the probability density  at $\vec{r}$.  We find that 
\begin{widetext}
\be
p(n(\vec{r}))dn(\vec{r})=\lim_{N\to\infty} \frac{A}{N}P(\gamma)d\gamma
\lim_{N\to\infty} \frac{A}{N}(N-1)[1-\frac{A}{N}n(\vec{r})]^N
(1-\frac{A}{N}n(\vec{r}))^{-3/2}dn(\vec{r})=Ae^{-An(\vec{r})}dn(\vec{r})\,.
\ee
\end{widetext}
We have thus shown that the distribution is exponential.  For a billiard system, 
the averaged density $n_0=1/A$ . Therefore, we have 
\be
p(n)=e^{-n/n_0}/n_0\,.
\ee

\subsection{Proof of Porter-Thomas distribution of eigenstate wave function}
For an eigenstate wave function,  everything is the same except that all the $\alpha_i$ 
are real rather than complex. So, we have $\sum_{i=1}^N \alpha_i^2=1$, and similarly
the distribution becomes
\be
P(\gamma)=\frac{\int d\alpha_1 \cdots d\alpha_N\delta(\gamma-\alpha_j^2)\delta(1-\sum_{i=1}^N\alpha_i^2)}{d\alpha_1 \cdots d\alpha_N \delta(1-\sum_{i=1}^N\alpha_i^2)}\,.
\ee
Let $\alpha_i^2=x$. The denominator equals the area of the $N$-dimensional hypersphere with unit radius. The numerator includes a factor of the area of a $(N-1)$-dimensional hypersphere with radius $\sqrt{1-x}$. This leads to 
\ba
P(\gamma)&=&\frac{\int_0^\infty dx x^{-1/2}\delta(\gamma-x)\frac{2\pi^{(N-1)/2}}{\Gamma((N-1)/2)}(\sqrt{1-x})^{N-2}}{2\pi^{N/2}/\Gamma(N/2)}\nonumber\\
&=&\sqrt{\frac{1}{\gamma\pi}}(\sqrt{1-\gamma})^{N-2}\frac{\Gamma(N/2)}{\Gamma((N-1)/2)}
\ea
With $\gamma=\frac{A}{N}n(\vec{r})$, we arrive at the Porter-Thomas distribution  in the limit of large $N$, 
\be
p(n)
=\sqrt{\frac{A}{2\pi n}}e^{-\frac{A}{2}n}
\ee
where we have used $\lim_{N\to\infty}\frac{\Gamma(N/2)}{\Gamma((N-1)/2)\sqrt{N}}=\frac{1}{\sqrt{2}}$.  

\end{document}